\documentclass[onecolumn,slac_one]{revtex4}
\usepackage{latexsym}
\usepackage{graphicx}
\usepackage{fancyhdr}
\renewcommand{\bar}[1]{\overline{#1}}

\newcommand{\ket}[1]{\,\left|\,{#1}\right\rangle}

\newcommand{\mbf}{\mathbf}

\begin{document}

\title{ADS/CFT and QCD}

\author{Stanley J. Brodsky}

\affiliation{Stanford Linear Accelerator Center, Stanford University, Stanford, California 94309}

\author{Guy F. de T\'eramond$^*$ }

\affiliation{Universidad de Costa Rica, San Jos\'e, Costa Rica,\\
and Stanford Linear Accelerator Center, Stanford University, Stanford, California 94309 \\
$^*$E-mail: gdt@asterix.crnet.cr}

\begin{abstract}
The AdS/CFT correspondence between string theory in AdS space and
conformal field theories in physical space-time leads to an
analytic, semi-classical model for strongly-coupled QCD which has
scale invariance and dimensional counting at short distances and
color confinement at large distances.  Although QCD is not
conformally invariant, one can nevertheless use the mathematical
representation of the conformal group in five-dimensional anti-de
Sitter space to construct a first approximation to the theory.  The
AdS/CFT correspondence also provides insights into the inherently
non-perturbative aspects of QCD, such as the orbital and radial
spectra of hadrons and the form of hadronic wavefunctions.  In
particular, we show that there is an exact correspondence between
the fifth-dimensional coordinate of AdS space $z$ and a specific
impact variable $\zeta$ which measures the separation of the quark
and gluonic constituents within the hadron in ordinary space-time.
This connection allows one to compute the analytic form of the
frame-independent light-front wavefunctions,
the fundamental entities which encode hadron properties and
allow the computation of decay constants, form factors, and other
exclusive scattering amplitudes. New
relativistic light-front equations in ordinary space-time are found
which reproduce the results obtained using the 5-dimensional
theory.  The effective light-front equations possess remarkable
algebraic structures and integrability properties.
Since they are complete and orthonormal, the AdS/CFT model
wavefunctions can also be used as a basis for the diagonalization of
the full light-front QCD Hamiltonian, thus systematically improving
the AdS/CFT approximation.
\end{abstract}

%
%\PACS{
%      {12.38.Aw,12.38.-t,11.25.Hf,11.90.+t,11.55.Jy}{}
%     } % end of PACS codes
%} %end of abstract
%\maketitle
%

\maketitle
\thispagestyle{fancy}

\section{The Conformal Approximation to QCD}

Quantum Chromodynamics, the Yang-Mills local gauge field theory of
$SU(3)_C$ color symmetry provides a fundamental understanding of
hadron and nuclear physics in terms of quark and gluon degrees of
freedom. However, because of its strong-coupling nature, it is
difficult to find analytic solutions to QCD or to make precise
predictions outside of its perturbative domain.  An important goal
is thus to find an initial approximation to QCD which is both
analytically tractable and which can be systematically improved. For
example, in quantum electrodynamics, the  Schr\"odinger and
Dirac equations provide accurate first approximations to
atomic bound state problems which can then be systematically
improved by using the Bethe-Salpeter formalism and correcting for
quantum fluctuations, such as the Lamb Shift and vacuum
polarization.

One of the most significant theoretical advances in recent years has
been the application of the AdS/CFT
correspondence~\cite{Maldacena:1997re} between string states defined
on the 5-dimensional Anti--de Sitter (AdS) space-time and conformal
field theories in physical space-time. The essential principle
underlying the AdS/CFT approach to conformal gauge theories is the
isomorphism of the group of Poincare' and conformal transformations
$SO(4,2)$ to the group of isometries of Anti-de Sitter space.  The
AdS metric is
\[ds^2 = \frac{R^2}{z^2}(\eta_{\mu \nu} dx^\mu
 dx^\nu - dz^2),\]
which is invariant under scale changes of the
coordinate in the fifth dimension $z \to \lambda z$ and $ x_\mu \to
\lambda x_\mu$.  Thus one can match scale transformations of the
theory in $3+1$ physical space-time to scale transformations in the
fifth dimension $z$.

QCD is not itself a conformal theory; however in the domain where
the QCD coupling is approximately constant and quark masses can be
neglected, QCD resembles a strongly-coupled conformal theory.
As shown by Polchinski
and Strassler~\cite{Polchinski:2001tt}, one can simulate confinement
by imposing boundary conditions in the holographic variable at
$z = z_0 ={1/\Lambda_{QCD}}$.
Confinement can also be
introduced by modifying the AdS metric to mimic a confining
potential. The resulting models, although {\it ad hoc}, provide a
simple semi-classical approximation to QCD which has both
counting-rule behavior~\cite{Brodsky:1973kr,Brodsky:1974vy,Matveev:1973ra,Lepage:1980fj}
at short distances and confinement at large distances.  This simple approach, which has been described as a
``bottom-up" approach,  has been successful
in obtaining general properties of scattering amplitudes  of
hadronic bound states~\cite{Janik:1999zk,Polchinski:2001tt,Polchinski:2002jw,Brower:2002er,Giddings:2002cd,Brodsky:2003px,Andreev:2004sy} and the
low-lying hadron spectra~\cite{Boschi-Filho:2002vd,deTeramond:2004qd,deTeramond:2005su,Erlich:2005qh,Katz:2005ir,Karch:2006pv,Brodsky:2006uq, Hong:2006ta}.
Studies of hadron couplings and chiral symmetry
breaking~\cite{Erlich:2005qh,Hong:2004sa,DaRold:2005zs,Hirn:2005vk,Ghoroku:2005vt},
quark potentials in confining
backgrounds~\cite{Boschi-Filho:2005mw,Andreev:2006ct} and  pomeron
physics~\cite{Boschi-Filho:2005yh,Brower:2006ea} has also been addressed
within the bottom-up approach to
holographic QCD, also known as AdS/QCD.

Recently, the behavior of the space-like form factors
of the pion~\cite{Radyushkin:2006iz,Brodsky:2006ys,Choi:2006ha}
and nucleons~\cite{deTeramond:2006xb} has been discussed within the framework of AdS/QCD.
Studies of geometric back-reaction controlling the
infrared physics are given in refs.~\cite{Csaki:2006ji,Shock:2006gt}.
It is also remarkable that the dynamical properties of
the quark-gluon plasma observed at RHIC~\cite{Muller:2006ee}
can be computed within the AdS/CFT
correspondence~\cite{Policastro:2001yc}.
In contrast to the simple bottom-up approach described above,
the introduction of additional higher
dimensional branes to the ${\rm AdS}_5 \times {\rm S}^5$
background has been used to study chiral symmetry breaking~\cite{Babington:2003vm},
and recently baryonic properties by using D4-D8 brane constructs~\cite{Nawa:2006gv,Hong:2007kx,Hata:2007mb}.

It was originally believed that the AdS/CFT mathematical tool would
only be applicable to strictly conformal theories such as
$\mathcal{N}=4$ supersymmetry.   In our approach, we will apply AdS/CFT
to the low momentum, strong coupling regime of QCD where the coupling is
approximately constant.
Theoretical~\cite{Alkofer:2004it} and phenomenological~\cite{Brodsky:2002nb}
evidence is in fact accumulating that the QCD couplings defined from physical
observables such as $\tau$ decay~\cite{Brodsky:1998ua} become constant at small virtuality; {\em i.e.}, effective charges develop an infrared fixed point in contradiction to the usual assumption of singular growth in the infrared.  Recent lattice gauge theory simulations~\cite{Furui:2006py}  also indicate an infrared fixed point for QCD.

It is clear from a physical perspective that in a confining theory where gluons and quarks have an effective mass or maximal wavelength, all
vacuum polarization corrections to the gluon self-energy must decouple at long wavelength; thus an infrared fixed point appears to be a natural consequence of confinement. Furthermore, if one considers a semi-classical approximation to QCD with massless quarks and without
particle creation or absorption, then the resulting $\beta$ function is zero, the coupling is constant, and the approximate theory is
scale and conformal invariant.

In the case of hard exclusive reactions~\cite{Lepage:1980fj}, the virtuality of the gluons exchanged in the underlying QCD process is typically much less than the momentum transfer scale $Q$ since typically several gluons share the total momentum transfer.  Since the coupling is probed in the conformal window, this kinematic feature can explain why the measured  proton Dirac form factor scales as $Q^4 F_1(Q^2) \simeq {\rm const}$ up to $Q^2 < 35$ GeV$^2$~\cite{Diehl:2004cx} with little sign of the logarithmic running of the QCD coupling.

One can  also use conformal symmetry as a {\it template}~\cite{Brodsky:1999gm},
systematically correcting for its nonzero $\beta$ function as well
as higher-twist effects. For example, ``commensurate scale
relations"~\cite{Brodsky:1994eh} which relate QCD observables to each
other, such as the generalized Crewther
relation~\cite{Brodsky:1995tb}, have no renormalization scale or
scheme ambiguity and retain a convergent perturbative structure
which reflects the underlying conformal symmetry of the classical
theory.  In general, the scale is set such that one has the correct
analytic behavior at the heavy particle
thresholds~\cite{Brodsky:1982gc}.
The importance of using an analytic
effective charge~\cite{Brodsky:1998mf} such as the pinch
scheme~\cite{Binger:2006sj,Cornwall:1989gv} for unifying the
electroweak and strong couplings and forces is also
important~\cite{Binger:2003by}.  Thus conformal symmetry is a useful
first approximant even for physical QCD.

In the AdS/CFT duality, the amplitude $\Phi(z)$ represents the
extension of the hadron into the compact fifth dimension.  The
behavior of $\Phi(z) \to z^\Delta$ at $z \to 0$ must match the
twist-dimension of the hadron at short distances $x^2 \to 0$. As we
shall discuss, one can use holography to map the amplitude
$\Phi(z)$ describing the hadronic state in the fifth dimension of
Anti-de Sitter space $\rm{AdS}_5$  to the light-front wavefunctions
$\psi_{n/h}$ of hadrons in physical
space-time~\cite{Brodsky:2006uq},  thus providing a relativistic
description of hadrons in QCD at the amplitude level.
In fact, there is an exact correspondence between the
fifth-dimensional coordinate of anti-de Sitter space $z$ and a
specific impact variable $\zeta$ in the light-front formalism which
measures the separation of the constituents within the hadron in
ordinary space-time.  We derive
this correspondence by noticing that the mapping of $z \to \zeta$
analytically transforms the expression for the form factors in
AdS/CFT to the exact QCD Drell-Yan-West expression in terms of
light-front wavefunctions.

Light-front wavefunctions are relativistic and frame-independent
generalizations of the familiar Schr\"odinger wavefunctions of
atomic physics, but they are determined at fixed light-cone time
$\tau= t +z/c$---the ``front form" advocated by Dirac---rather than
at fixed ordinary time $t$.  An important advantage of light-front
quantization is the fact that it provides exact formulas to write
down matrix elements as a sum of bilinear forms, which can be mapped
into their AdS/CFT counterparts in the semi-classical approximation.
One can thus obtain not only an
accurate description of the hadron spectrum for light quarks, but also
a remarkably simple but realistic model of the valence wavefunctions
of mesons, baryons, and glueballs.
The light-front wavefunctions predicted by AdS/QCD have many phenomenological applications ranging from exclusive $B$ and $D$ decays, deeply virtual Compton scattering and exclusive reactions such as form factors, two-photon processes, and two-body scattering.
One thus obtains a connection between the theories and tools used in string theory and the fundamental phenomenology  of hadrons.

\section{Light-Front Wavefunctions in Impact Space}

The light-front expansion is constructed by quantizing QCD
at fixed light-cone time~\cite{Dirac:1949cp} $\tau = t + z/c$ and
forming the invariant light-front Hamiltonian: $ H^{QCD}_{LF} = P^+
P^- - {\vec P_\perp}^2$ where $P^\pm = P^0 \pm
P^z$~\cite{Brodsky:1997de}.
The momentum generators $P^+$ and $\vec
P_\perp$ are kinematical; {\em i.e.}, they are independent of the
interactions. The generator $P^- = i {d\over d\tau}$ generates
light-cone time translations, and the eigen-spectrum of the Lorentz
scalar $ H^{QCD}_{LF}$ gives the mass spectrum of the color-singlet
hadron states in QCD; the projection of the eigensolution on the free Fock basis gives the hadronic light-front
wavefunctions.

The holographic mapping of hadronic LFWFs to AdS string modes
is most transparent when one uses the impact parameter space representation~\cite{Soper:1976jc}.
The total position coordinate of a hadron or its transverse center
of momentum $\mbf{R}_\perp$, is defined in terms of the energy
momentum tensor $T^{\mu \nu}$
\begin{equation}
\mbf{R}_\perp = \frac{1}{P^+} \int dx^-
\negthinspace \int d^2 \mbf{x}_\perp \,T^{++} \,
\mbf{x}_\perp.
\end{equation}
In terms of partonic transverse coordinates
\begin{equation}
x_i \mbf{r}_{\perp i} = x_i \mbf{R}_\perp + \mbf{b}_{\perp i},
\end{equation}
where  the $\mbf{r}_{\perp i}$ are the physical transverse position
coordinates and  $\mbf{b}_{\perp i}$ frame independent  internal
coordinates, conjugate to the relative coordinates $\mbf{k}_{\perp i}$.
Thus, $\sum_i \mbf{b}_{\perp i} = 0$ and
$\mbf{R}_\perp = \sum_i x_i \mbf{r}_{\perp i}$.
The LFWF $\psi_n(x_j, \mbf{k}_{\perp j})$ can be expanded in terms of the $n-1$ independent
coordinates $\mbf{b}_{\perp j}$,  $j = 1,2,\dots,n-1$
\begin{equation} \label{eq:LFWFb}
\psi_n(x_j, \mbf{k}_{\perp j}) =  (4 \pi)^\frac{(n-1)}{2}
\prod_{j=1}^{n-1}\int d^2 \mbf{b}_{\perp j}
\exp\!{\Big(i \sum_{j=1}^{n-1} \mbf{b}_{\perp j} \cdot \mbf{k}_{\perp j}\Big)}
\widetilde{\psi}_n(x_j, \mathbf{b}_{\perp j}).
\end{equation}
The normalization is defined by
\begin{equation}
\sum_n  \prod_{j=1}^{n-1} \int d x_j d^2 \mbf{b}_{\perp j}
\left\vert\widetilde \psi_n(x_j, \mbf{b}_{\perp j})\right\vert^2 = 1.
\end{equation}

One of the important advantages of the light-front formalism is that current
matrix elements can be represented without approximation as overlaps
of light-front wavefunctions. In the case of the elastic space-like
form factors, the matrix element of the $J^+$ current only couples
Fock states with the same number of constituents.
If
the charged parton $n$ is the active constituent struck by the
current, and the quarks $i = 1,2, \dots ,n-1$ are spectators, then
the Drell-Yan West formula~\cite{Drell:1969km,West:1970av,Brodsky:1980zm} in impact space is
\begin{equation} \label{eq:FFb}
F(q^2) =  \sum_n  \prod_{j=1}^{n-1}\int d x_j d^2 \mbf{b}_{\perp j}
\exp\!{\Bigl(i \mbf{q}_\perp \cdot \sum_{j=1}^{n-1} x_j \mbf{b}_{\perp j}\Bigr)}
\left\vert \widetilde \psi_n(x_j, \mbf{b}_{\perp j})\right\vert^2,
\end{equation}
corresponding to a change of transverse momenta $x_j \mbf{q}_\perp$ for each
of the $n-1$ spectators.  This is a convenient form
for comparison with AdS results, since the form factor
is expressed in terms of the product of light-front wave
functions  with identical variables.

We can now establish an explicit connection between the AdS/CFT and the LF formulae.
It is useful to express (\ref{eq:FFb}) in terms of an effective single
particle transverse distribution $\widetilde \rho$ ~\cite{Brodsky:2006uq}
\begin{equation} \label{eq:FFzeta}
F(q^2) = 2 \pi \int_0^1 dx \frac{(1-x)}{x}  \int \zeta d \zeta\,
J_0\negthinspace\left(\zeta q \sqrt{\frac{1-x}{x}}\right) \tilde \rho(x,\zeta),
\end{equation}
where we have introduced the variable
\begin{equation}
\zeta = \sqrt{\frac{x}{1-x}} ~\Big\vert \sum_{j=1}^{n-1} x_j \mathbf{b}_{\perp j}\Big\vert,
\end{equation}
representing the $x$-weighted transverse impact coordinate of the
spectator system. On the other hand, the expression for the form factor in AdS
space is represented as the overlap in the fifth dimension coordinate $z$
of the normalizable modes dual to the incoming
and outgoing hadrons, $\Phi_P$ and $\Phi_{P'}$, with the
non-normalizable mode, $J(Q,z) = z Q K_1(z Q)$, dual to the external
source~\cite{Polchinski:2002jw}
\begin{equation}
F(Q^2)
= R^{3} \int \frac{dz}{z^{3}} \,
  \Phi_{P'}(z) J(Q,z) \Phi_P(z).
\label{eq:FFAdS}
\end{equation}
If we compare (\ref{eq:FFzeta}) in impact space with the expression for the form factor in
AdS space (\ref{eq:FFAdS}) for arbitrary values of $Q$ using the identity
\begin{equation} \label{eq:intJ}
\int_0^1 dx \, J_0\negthinspace\left(\zeta Q
\sqrt{\frac{1-x}{x}}\right) = \zeta Q K_1(\zeta Q),
\end{equation}
then we can identify the spectator density
function appearing in the light-front
formalism with the corresponding AdS density
\begin{equation} \label{eq:hc}
\tilde \rho(x,\zeta)
 =  \frac{R^3}{2 \pi} \frac{x}{1-x}
\frac{\left\vert \Phi(\zeta)\right\vert^2}{\zeta^4}.
\end{equation}
Equation (\ref{eq:hc}) gives a precise relation between  string modes $\Phi(\zeta)$
in AdS$_5$ and the QCD transverse charge density $\tilde\rho(x,\zeta)$.
The variable $\zeta$ represents a measure of the transverse
separation between point-like constituents, and it is also the
holographic variable $z$
characterizing the string scale in AdS. Consequently the AdS string
mode $\Phi(z)$ can be regarded as the propability amplitude to find $n$
partons at transverse impact separation $\zeta = z$. Furthermore, its
eigenmodes determine the hadronic spectrum~\cite{Brodsky:2006uq}.

In the case of a two-parton constituent bound state, the correspondence
between the string amplitude $\Phi(z)$ and the light-front wave
function $\widetilde\psi(x,\mathbf{b})$ is expressed in closed form\cite{Brodsky:2006uq}
\begin{equation}  \label{eq:Phipsi}
\left\vert\widetilde\psi(x,\zeta)\right\vert^2 =
\frac{R^3}{2 \pi} ~x(1-x)~
\frac{\left\vert \Phi(\zeta)\right\vert^2}{\zeta^4},
\end{equation}
where $\zeta^2 = x(1-x) \mathbf{b}_\perp^2$.
Here $b_\perp$ is the impact separation and Fourier conjugate to $k_\perp$.

\section{Holographic Light-Front Representation}

The equations of motion in AdS space can be
recast in the form of  a light-front Hamiltonian equation~\cite{Brodsky:1997de}
\begin{equation}
H_{LC} \ket{\psi_h} = \mathcal{M}^2 \ket{\psi_h},
\label{eq:HLC}
\end{equation}
a remarkable result which  allows the discussion of the AdS/CFT
solutions in terms of light-front equations in physical 3+1 space-time.
By substituting
$\phi(\zeta) = \left(\frac{\zeta}{R}\right)^{-3/2} \Phi(\zeta)$,
in the AdS wave equation describing the propagation of scalar modes in AdS space
\begin{equation} \label{eq:AdSPhiM}
\left[z^2 \partial_z^2 - (d-1) z\,\partial_z + z^2 \mathcal{M}^2 - (\mu R)^2 \right] \Phi(z) = 0,
\end{equation}
we find an effective Schr\"odinger equation as a function of the
weighted impact variable $\zeta$
\begin{equation} \label{eq:Scheq}
\left[-\frac{d^2}{d \zeta^2} + V(\zeta) \right] \phi(\zeta) = \mathcal{M}^2 \phi(\zeta),
\end{equation}
with the effective potential
$V(\zeta) \to - (1-4 L^2)/4\zeta^2$ in the conformal limit,
where we identity $\zeta$ with
the fifth dimension $z$ of AdS space: $\zeta = z$.
We have written above $(\mu R)^2 = - 4 + L^2$.  The solution
to (\ref{eq:Scheq}) is
$\phi(z) = z^{-\frac{3}{2}} \Phi(z) = C z^\frac{1}{2} J_L(z \mathcal{M}).$
This equation reproduces the AdS/CFT solutions for mesons with relative orbital
 angular momentum $L$.
The holographic hadronic light-front wave functions
$\phi(\zeta) = \langle \zeta \vert \psi_h \rangle$ are normalized
according to
\begin{equation}
\langle \psi_h \vert \psi_h \rangle =
\int d\zeta \, \vert \langle  \zeta \vert \psi_h \rangle \vert^2 = 1,
\end{equation}
and represent the probability
amplitude to find $n$-partons at transverse impact separation $\zeta =
z$. Its eigenmodes determine the hadronic mass spectrum.

\begin{figure}[ht]
\centering
\includegraphics[angle=0,width=10.6cm]{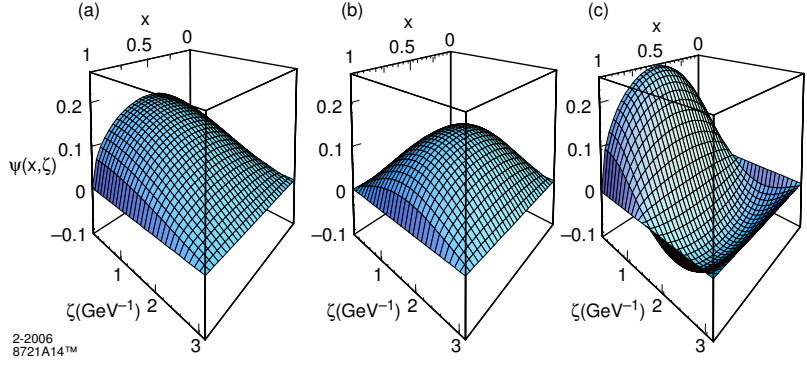}
\caption{AdS/QCD Predictions for the light-front wavefunctions of a meson.}
\label{fig:MesonLFWF}
\end{figure}

The lowest stable state $L = 0$ is determined by the Breitenlohner-Freedman
bound\cite{Breitenlohner:1982jf}.   Its
eigenvalues are set by the boundary conditions at $\phi(z = 1/\Lambda_{\rm QCD}) = 0$ and are
given in terms of the roots of  Bessel functions:
$\mathcal{M}_{L,k} = \beta_{L,k} \Lambda_{\rm QCD}$.
Normalized LFWFs $\widetilde\psi_{L,k}$ follow from
(\ref{eq:Phipsi})
\begin{equation}
\widetilde \psi_{L,k}(x, \zeta)
=  B_{L,k} \sqrt{x(1-x)}
J_L \left(\zeta \beta_{L,k} \Lambda_{\rm QCD}\right)
\theta\big(z \le \Lambda^{-1}_{\rm QCD}\big),
\end{equation}
where $B_{L,k} = {\Lambda_{\rm QCD}}/ \sqrt{ \pi} J_{1+L}(\beta_{L,k})$.
The resulting wavefunctions depicted in Fig. \ref{fig:MesonLFWF}
display confinement at large interquark
separation and conformal symmetry at short distances, reproducing dimensional counting rules for hard exclusive processes and the scaling and conformal properties of the LFWFs at high relative
momenta in agreement  with perturbative QCD.

Since they are complete and orthonormal, these AdS/CFT model
wavefunctions can be used as an initial ansatz for a variational
treatment or as a basis for the diagonalization of the light-front QCD Hamiltonian.  We are now in fact investigating this possibility with J. Vary and A. Harinandrath.
Alternatively one can introduce confinement by adding a harmonic oscillator potential
$\kappa^4 z^2$ to the conformal kernel in  Eq. (\ref{eq:Scheq}).
One can also introduce nonzero quark masses for the meson. The procedure is straightforward
in the $\mathbf{k}_\perp$ representation by using the substitution
$\frac{\mathbf{k}_\perp^2}{x(1-x)} \to \frac{\mathbf{k}_\perp^2 +  m^2_2}{x}
+ \frac{ \mathbf{k}_\perp^2 + m^2}{1-x}$.

\section{Integrability of AdS/CFT Equations}

The integrability methods of
Ref. [\cite{Infeld:1941}] find a remarkable application in the AdS/CFT
correspondence.   Integrability follows
if the equations describing a physical model can be factorized in
terms of linear operators. These ladder operators then
generate all the eigenfunctions once the lowest mass eigenfunction is known.
In holographic QCD, the conformally invariant 3 + 1 light-front differential equations can be expressed as ladder operators and their solutions can then be expressed in terms of analytical functions.

In the conformal limit the ladder algebra for bosonic ($B$) or fermionic ($F$) modes is given in terms of
the operator ($\Gamma^B =1, ~~ \Gamma^F = \gamma_5$)
\begin{equation} \label{eq:Pi}
\Pi_\nu^{B,F}(\zeta) = -i \left( \frac{d}{d \zeta} - \frac{\nu + {1\over 2}}{\zeta} \, \Gamma^{B,F} \!\right) ,
\end{equation}
and its adjoint
\begin{equation}
\Pi^{B,F}_\nu(\zeta)^\dagger = -i \left(\frac{d}{d \zeta} + \frac{\nu + {1\over 2}}{\zeta}
\, \Gamma^{B,F} \!\right) ,
\end{equation}
with commutation relations
\begin{equation}
\left[\Pi_\nu^{B,F}(\zeta),\Pi_\nu^{B,F}(\zeta)^\dagger\right]
=  \frac{2 \nu+1}{\zeta^2} \, \Gamma^{B,F} .
\end{equation}
For $\nu \ge 0$ the Hamiltonian is written as a bilinear form
$H^{B,F}_{LC} = {\Pi_\nu^{B,F}}^\dagger \Pi_\nu^{B,F}$. In the fermionic case the eigenmodes
also satisfy a first order LF Dirac equation.
For bosonic modes, the lowest stable
state $\nu =0$ corresponds to the Breitenlohner-Freedman bound. Higher orbital states
are constructed from the L-th application of the raising operator $a^\dagger = - i \Pi^B$ on the
ground state.

\section{Hadronic Spectra in AdS/QCD}

The holographic model
based on truncated AdS space can be used to obtain the hadronic
spectrum of light quark $q \bar q, qqq$ and $gg$ bound states.  Specific
hadrons are identified by the correspondence of the amplitude in the fifth dimension with
the twist dimension of the interpolating operator for the hadron's valence
Fock state, including its orbital angular momentum excitations.   Bosonic modes with conformal
dimension $2+L$ are dual to the interpolating operator $\mathcal{O}_{\tau + L}$ with $\tau = 2$.
For fermionic modes $\tau = 3$.
For example,
the set of three-quark baryons with spin 1/2 and higher  is described
by the light-front Dirac equation
\begin{equation}
\left(\alpha \, \Pi^F \! (\zeta) - \mathcal{M} \right)  \psi(\zeta) = 0,
\end{equation}
where
$i \alpha =
\left( \begin{array}{cc}
0 & I \\
-I & 0
\end{array} \right)$
in the Weyl representation.
The solution is\begin{equation} \label{eq:DiracLF}
\psi(\zeta) =  C \sqrt{\zeta}
\left[J_{L+1}\left(\zeta \mathcal{M} \right) \, u_+
+ J_{L+2}\left(z \mathcal{M}\right) \, u_- \right],
\end{equation}
with $\gamma_5 u_\pm = u_\pm$.
A discrete  four-dimensional spectrum follows when we impose the boundary condition
$\psi_\pm(\zeta=1/\Lambda_{\rm QCD}) = 0$:
$\mathcal{M}_{\alpha, k}^+ = \beta_{\alpha,k} \Lambda_{\rm QCD},
\mathcal{M}_{\alpha, k}^- = \beta_{\alpha + 1,k} \Lambda_{\rm QCD}$,
with a scale-independent mass ratio\cite{deTeramond:2005su}.

\begin{figure}[ht]
\centering
\includegraphics[angle=0,width=10.6cm]{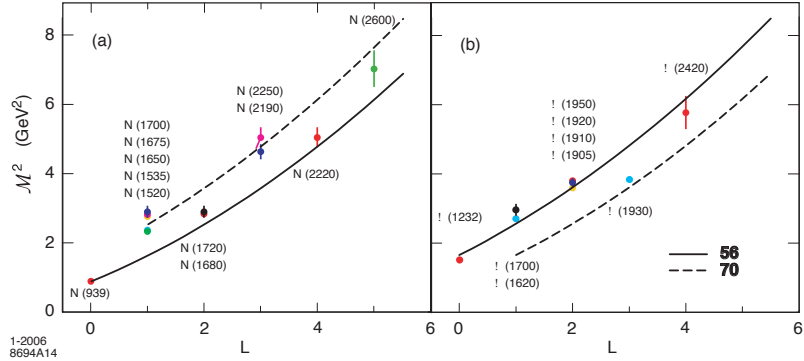}
\caption{Predictions for the light baryon orbital spectrum for
$\Lambda_{QCD}$ = 0.25 GeV. The  $\bf 56$ trajectory corresponds to
$L$ even  $P=+$ states, and the $\bf 70$ to $L$ odd  $P=-$ states.}
\label{fig:BaryonSpec}
\end{figure}

Figure \ref{fig:BaryonSpec}(a) shows the predicted orbital spectrum of the
nucleon states and Fig. \ref{fig:BaryonSpec}(b) the $\Delta$ orbital
resonances. The spin-3/2 trajectories are determined from the corresponding Rarita-Schwinger  equation.  The solution of the spin-3/2 for polarization along Minkowski coordinates, $\psi_\mu$, is similar to the spin-1/2 solution.
The data for the baryon spectra are from [\cite{Eidelman:2004wy}].
The internal parity of states is determined from the SU(6)
spin-flavor symmetry.
Since only
one parameter, the QCD mass scale $\Lambda_{QCD}$, is introduced, the
agreement with the pattern of physical states is remarkable. In
particular, the ratio of $\Delta$ to nucleon trajectories is
determined by the ratio of zeros of Bessel functions.  The
predicted mass spectrum in the truncated space model is linear $M \propto L$ at high orbital
angular momentum, in contrast to the quadratic dependence $M^2
\propto L$ in the usual Regge parameterization. One can obtain $M^2 \propto (L+n)$ dependence
in the holographic model by the introduction of a harmonic potential $\kappa^2 z^2$
in the AdS wave equations~\cite{Karch:2006pv}.  This result can also be obtained by extending the conformal
algebra written above.
An account of the extended algebraic holographic model and a possible supersymmetric
connection between the bosonic and fermionic operators used in the holographic
construction will be described elsewhere.

\section{Pion Form Factor}

\begin{figure}[h]
\centering
\includegraphics[angle=0,width=6.0cm]{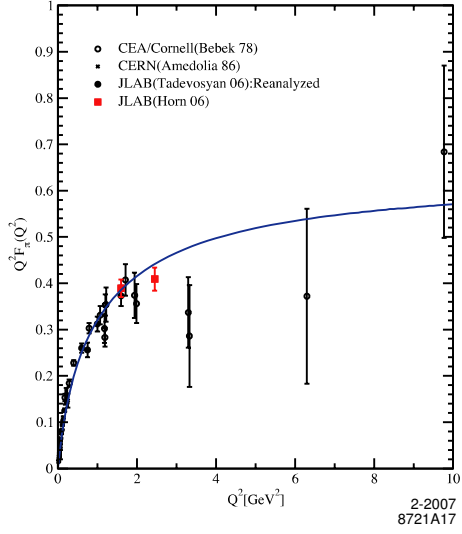}
\caption{$Q^2 F_\pi(Q^2)$ in the harmonic oscillator model
for $\kappa =  0.4$ GeV.}
\label{fig:pionFF}
\end{figure}

Hadron form factors can be predicted from the overlap integral representation in AdS space or equivalently by using the Drell-Yan West formula in physical space-time.
For the pion string mode $\Phi$ in the harmonic oscillator model~\cite{Karch:2006pv}
\begin{equation} \label{eq:TOPM}
\Phi_\pi^{HO}(z) = \frac{\sqrt{2} \kappa}{R^{3/2}}\, z^2 \, e^{-\kappa^2 z^2/2},
\end{equation}
the form factor has a closed form solution
\begin{equation}
F(Q^2) = 1 + \frac{Q^2}{4 \kappa^2} \exp\left(\frac{Q^2}{4\kappa^2}\right)
 Ei\left(-\frac{Q^2}{4\kappa^2}\right),
\end{equation}
where $Ei$ is the exponential integral
\begin{equation}
Ei(-x) = \int_\infty^x e^{-t} \frac{dt}{t}.
\end{equation}
Expanding the function $Ei(-x)$ for large arguments, we find for  $- Q^2 \gg \kappa^2$
\begin{equation}
F(Q^2) \to \frac{4 \kappa^2}{Q^2},
\end{equation}
and we recover the dimensional counting rule.
The prediction for the pion form factor is
shown in Fig.~\ref{fig:pionFF}. The space-like behavior of the  pion
form factor in the harmonic oscillator (HO)
model  is almost indistinguishable from the
truncated-space (TS) model result.
The form factor at high $Q^2$
receives contributions from small $\zeta$, corresponding to small
$\vec b_\perp \sim {\cal O}(1/Q)$ (high relative $\vec k_\perp \sim
 {\cal O}(Q)$),
as well as $x \to 1$.  The AdS/CFT dynamics is thus distinct
from endpoint models~\cite{Radyushkin:2006iz} in which the LFWF is
evaluated solely at small transverse momentum or large impact
separation.

The $x \to 1$ endpoint domain is often referred to as a ``soft''
Feynman contribution. In fact $x \to 1$ for the struck quark
requires that all of the spectators have $x = k^+/P^+ = (k^0+
k^z)/P^+  \to 0$; this  in turn requires high longitudinal momenta
$k^z \to - \infty$ for all spectators  --  unless one has both
massless spectator quarks $m \equiv 0$ with zero transverse momentum
$k_\perp \equiv 0$, which is a regime of measure zero. If one uses a
covariant formalism, such as the Bethe-Salpeter theory, then the
virtuality of the struck quark  becomes  infinitely spacelike:
$k^2_F \sim  - \frac{k^2_\perp + m^2}{ 1-x}$  in the endpoint domain.
Thus, actually,  $x \to 1$ corresponds to high relative longitudinal
momentum; it is as hard a domain in the hadron wavefunction as high
transverse momentum.

\section{The Pion Decay Constant}

The pion decay constant is given by the matrix element of the axial isospin
current $J^{\mu 5 a}$ between a physical pion and the vacuum state~\cite{Peskin:1995ev}
$\bigl\langle 0 \bigl\vert J_W^+(0) \bigr\vert \pi^-(P^+, \vec P_\perp) \bigr\rangle$,
where $J_W^+$ is the flavor changing weak current.
Only the valence state
with $L_z = 0$, $S_z = 0$, contributes to the decay of the $\pi^\pm$.
Expanding the  hadronic initial state in the decay amplitude
into its Fock components we find
 \begin{equation}
f_\pi =  2 \sqrt{N_C} \int_0^1 dx \int \frac{d^2 \vec k_\perp}{16 \pi^3}
~\psi_{\bar q q/\pi}(x,k_\perp).
\end{equation}
This light-cone equation allows the exact computation of the pion decay
constant in terms of the valence pion light-front wave function~\cite{Lepage:1980fj}.

The meson distribution amplitude $\phi(x,Q)$ is defined as~\cite{Lepage:1979zb}
\begin{equation}
\phi(x,Q) = \int^{Q^2} \frac{d^2 \vec k_\perp}{16 \pi^3}~\psi(x, k_\perp).
\end{equation}
It follows that
\begin{equation} \label{eq:phix}
\phi_\pi(x, Q\to\infty)
= \frac{4}{\sqrt{3}\pi}  f_\pi \sqrt{x(1-x)},
\end{equation}
with
\begin{equation}
f_\pi = \frac{1}{8} \sqrt{\frac{3}{2}} \, R^{3/2} \lim_{\zeta \to 0}
\frac{\Phi(\zeta)}{\zeta^2},
\end{equation}
since $\phi(x, Q \to \infty) \to
\widetilde \psi(x,\vec b_\perp \to 0)/\sqrt{4 \pi}$ and
$\Phi_\pi \sim \zeta^2$ as $\zeta\to 0$.
The pion decay constant depends only on the behavior of the AdS
string mode near the asympototic boundary, $\zeta = z = 0$ and the
mode normalization. For the truncated-space (TS) pion mode we find
$
f_\pi^{TS} = \frac{\sqrt{3}}{8 J_1(\beta_{0,k})} \, \Lambda_{\rm QCD}
= 83. 4 ~{\rm Mev},$
for $\Lambda_{QCD} = 0.2$ MeV. The corresponding result for the
transverse harmonic oscillator (HO) pion mode (\ref{eq:TOPM} ) is
$
f_\pi^{HO} =  \frac{\sqrt{3}}{8} \, \kappa = 86.6 ~{\rm MeV},
$
for $\kappa = 0.4$ GeV.  The values of $\Lambda_{QCD}$ and $\kappa$ are
determined from the space-like form factor data as discussed above.
The experimental result for $f_\pi$ is extracted form the rate
of weak $\pi$ decay and has the value
$f_\pi = 92.4$ MeV~\cite{Eidelman:2004wy}.

It is interesting to note that the pion distribution amplitude
predicted by AdS/QCD (\ref{eq:phix})
has a quite different $x$-behavior than the
asymptotic distribution amplitude predicted from the PQCD
evolution~\cite{Lepage:1979zb} of the pion distribution amplitude
$\phi_\pi(x,Q \to \infty)= \sqrt 3  f_\pi x(1-x) $.  The broader
shape of the pion distribution increases the magnitude of the
leading twist perturbative QCD prediction for the pion form factor
by a factor of $16/9$ compared to the prediction based on the
asymptotic form, bringing the PQCD prediction  close to the
empirical pion form factor~\cite{Choi:2006ha}.

\vspace{10pt}

\noindent{\bf Acknowledgments}

\vspace{5pt}

This research was supported by the Department
of Energy contract DE--AC02--76SF00515. We thank Alexander Gorsky, Chueng-Ryong Ji,  and Mitat Unsal
for helpful comments.
\bigskip

%\noindent{\bf Acknowledgments}
%\vspace{4pt}Work supported by the Department of Energy under
%contract number DE--AC02--76SF00515. The AdS/CFT results reported
%here were done in collaboration with Guy de T\'eramond.
%\bigskip
%\bigskip
%\bigskip
%\bigskip

\end{document}